\documentclass[aps,preprint,showpacs,floats,epsf,epsfig,nofootinbib,12pt]{revtex4}
\usepackage{graphicx}
\usepackage{epsfig}
\usepackage[dvips]{color}
\usepackage{subfigure}

\def\beq{\begin{eqnarray}}
\def\eeq{\end{eqnarray}}
\def\non{\nonumber}

\begin{document}

\title{A natural interpretation on the data of $\Lambda_c\to\Sigma\pi$}

\vspace{1cm}

\author{ Hong-Wei Ke$^{1}$\footnote{khw020056@tju.edu.cn}
and Xue-Qian Li$^2$\footnote{lixq@nankai.edu.cn} }

\affiliation{  $^{1}$ School of Science, Tianjin University,
Tianjin 300072, China
\\
  $^{2}$ School of Physics, Nankai University, Tianjin 300071, China }

\vspace{12cm}

\begin{abstract}

Even though the Standard Model (SM) has achieved great success, its application to the field of low energies still lacks
solid foundation due to our limited knowledge on non-perturbative QCD.
Practically, all theoretical calculations of
the hadronic transition matrix elements are based various phenomenological models.
There indeed exist some anomalies in the field which are waiting
for interpretations. The goal of this work is trying to solve
one of the anomalies: the discrepancy between the theoretical prediction on the sign of the up-down asymmetry
parameter of $\Lambda_c\to\Sigma\pi$ and the experimental measurement.
In the literatures several authors calculated the rate and
determined the asymmetry
parameter within various schemes, but there exist obvious loopholes in those adopted scenarios. To solve the discrepancy between theory and data,
we suggest that
not only the direct transition process contributes to the observed $\Lambda_c\to\Sigma\pi$, but also other portals such as $\Lambda_c\to \Lambda\rho$
also play a substantial role via an isospin-conserving re-scattering $\Lambda\rho\to\Sigma\pi$. Taking into account
of the effects induced by the final state interaction, we re-evaluate the relevant quantities. Our numerical results indicate
that the new theoretical prediction based on this scenario involving an interference between the direct transition of $\Lambda_c\to\Sigma\pi$
and the portal $\Lambda_c\to\Lambda\rho\to\Sigma\pi$ can make both the decay rate and sign of the asymmetry parameter to be consistent with
data.

\pacs{13.30.-a,14.20.Lq, 12.40.-y}

\end{abstract}

\maketitle

\section{Introduction}

Even though the Standard Model (SM) has achieved great success, its application to the field of low energies still lacks
solid foundation due to our limited knowledge on non-perturbative QCD.
Practically, all theoretical calculations of
the hadronic transition matrix elements are based various phenomenological models. There
indeed exist some anomalies in the field which are waiting
for interpretations.
One of the anomalies is the discrepancy between the theoretical prediction on the sign of the up-down asymmetry
parameter of $\Lambda_c\to\Sigma\pi$ and the experimental measurement. In fact, besides the meson case, for
baryons which contain three ingredients, the complexity makes a thorough study on them more difficult than on
mesons.  However from another aspect, the involved physics in the transitions between baryons is also richer and
by the research one can get better understanding of the governing mechanisms. An advantage of studying decays of baryons involving heavy flavors
is obvious just as one does on the heavy mesons (structure, production and decay).

Especially, the charmed hadrons are of special significance because charm quark is heavier than the light quarks ($u,d,s$), but
at the same time is not as heavy as the bottom quark, so that relativistic effects are not negligible at all. The issue
that the lifetimes of $B^{\pm}, B^0 $ and $\Lambda_b$ are close, however, the lifetimes of $D^{\pm}, D^0$
and $\Lambda_c$ are quite apart, has been warmly discussed. It is believed that the Pauli interference induces the lifetime differences of $D^{\pm}$ and $D^0$
\cite{Bilic:1984nq,Bellini:1996ra} which is suppressed for the $B$-hadrons, but for $\Lambda_c$ the question still exists.

Since 1990s, many decay channels
of $\Lambda_c$ have successively been measured by  experimental collaborations\cite{Albrecht:1991bu,Avery:1990ya,Zupanc:2013iki,Ablikim:2015flg},
and the field  has attracted attentions of theorists. Its weak decays have been carefully explored with different approaches\cite{Korner:1992wi,Xu:1992vc,Cheng:1991sn,Cheng:1993gf,Sharma:1998rd,Cheng:1995fe,Cheng:2018hwl,Ivanov:1997ra,Zenczykowski:1993jm}.
In this work, we would re-visit the old topic because much larger data-sets with higher precision are available at BESIII, Belle and even LHCb which make
us to hope
getting better understanding on the charmed baryons.

Among the previous theoretical studies on the decay rate of $\Lambda_c\to\Sigma\pi$ and the corresponding up-down asymmetry parameter $\alpha$, the pole model
was adopted because of its advantage. The pole model is simple and the relevant parameters are adopted by fitting data, therefore one can trust its effectiveness.
A naive
conjecture would expect that the prediction obtained with this model should be consistent with data even though an error is unavoidable  in
this relatively rough picture. However, it is noticed that the prediction on the up-down asymmetry parameter $\alpha$ is positive while the measured value is negative.
This apparent discrepancy which is not a tolerable deviation, indicates that there must be something wrong.
Thus to solve this ``anomaly" there are two routes. One is that the method adopted for the calculation should be modified whereas another possibility is that besides
the direct transition, there exist other contributions to the observed data on $\Lambda_c\to\Sigma\pi$. As the first route,
Cheng and his collaborators \cite{Cheng:1993gf} went on to use the current algebra
calculating the transition matrix element $\langle \Sigma| H_{eff}|\Lambda_c\rangle$ and obtained negative $\alpha$. However, the pre-condition of
using the current algebra\cite{Weinberg:1966fm} is properly extracting the pion field out from the matrix element $\langle \pi\Sigma|H'_{eff}  |\Lambda_c\rangle$ under the
soft-pion approximation. By contrast for the process $\Lambda_c\to\Sigma\pi$, the 3-momentum of the pion is not small to be ``soft", thus
the whole scenario is questionable.

Instead, we follow the second route i.e. the obvious loophole may suggest that there could be another mechanism.
We propose that other channels of $\Lambda_c$ decays would contribute to the observed $\Lambda_c\to\Sigma\pi$ via final state interactions.
The interference between the direct transition and the new contribution may lead to the results consistent with data.
Considering the effective interaction and the decay rate of $\Lambda_c$ decays, the most possible
two-step process is that $\Lambda_c$ first transits into $\Lambda \rho$ then, by a re-scattering
$\Lambda\rho$ turns into $\Sigma\pi$ and its contribution would interfere with the amplitude of direct transition $\Lambda_c\to\Sigma\pi$.

As is well known, the weak decays of heavy hadrons mainly occur
via an emission of virtual $W$ or $Z$ bosons  which later turn into lepton or quark pairs, from the heavy quark(antiquark). In our case,
$\Lambda_c$ is an iso-spin singlet, therefore the $u$-$d$ subsystem (one may call it as a diquark) in $\Lambda_c$ exists in an iso-spin singlet ($I=0$),
and it is noted that the $u$-$d$ subsystem in
$\Lambda$ is also an isospin singlet, but in $\Sigma^0$ the subsystem is an isospin-triplet ($I=1$). During the transition of $\Lambda_c\to\Lambda$
the $u$-$d$ subsystem serves generally as a spectator and retains its isospin unchanged, whereas for $\Lambda_c\to\Sigma^0$, the isospin of the $u$-$d$  subsystem
is forced to change from a singlet into a triplet. Practically, for the $\Lambda_c\to\Sigma^0\pi^+$ transition,
the $W$-boson emitted by the charm quark must be connected to the $u$-$d$ subsystem which is no longer a real spectator.
Although the weak interaction
does not conserve isospin, the isospin analysis may help us to get an insight into what happens during the transition.

According to the analysis given in literature \cite{Cheng:1991sn},
the rate of $\Lambda_c\to \Lambda\rho$  should be about twice larger than that of $\Lambda_c\to \Sigma\pi$,
therefore
a two-steps process: $\Lambda_c\to\Lambda\rho\to\Sigma\pi$ may substantiate and change the picture (especially the sign of the asymmetry parameter).
This is not a surprise to notice the role of final interaction, in our earlier work,
we studied the case of $D^0\to K^0\bar K^0$ which has the same rate as $D^0\to K^+K^-$. In fact the former is strongly suppressed, nevertheless  the later is
favored. The result is fully understood as $D^0\to K^0\bar K^0$ is realized via a re-scattering $K^+K^-\to K^0\bar K^0$\cite{Dai:1999cs}.
Therefore in this work we include the contributions from both the direct transition and that induced by the final state interaction, and their
interference leads to the final result which is experimentally measured. For a comparison, in the following table, we list the results given in literature.

Concretely, all the coupled channels of $\Lambda_c$  should contribute  to the observed
$\Lambda_c\to \Sigma\pi$ via final state interactions (re-scattering). In terms of the effective interaction,
the corresponding coupling constants  the first step  of the sequential
$\Lambda_c$ decays could be formulated according to the scenario presented in the literature.
The re-scattering mechanism has been  successfully applied to  explain some
anomalies existing in low energy experiments, such as  the decays of $\Upsilon$ and bottomed
mesons\cite{Meng:2008dd,Ke:2010aw,Yuan:2012zw,Cheng:2004ru}, thus we have a full confidence
that the mechanism also works well  here.

This paper is organized as follows: after this introduction we
will consider the contribution to $\Lambda_c\to \Sigma\pi$ by including the final state interactions. In section III  we present our numerical results. Section IV is
devoted to a brief summary.

\begin{table}
\caption{Decay width $\Gamma$ (in units of $\times 10^{-14}$) and up-down asymmetry $\alpha$ of $\Lambda_c\to \Sigma \pi$. }\label{Tab:10}
\begin{ruledtabular}
\begin{tabular}{ccccccccc}
  &Ref.\cite{Korner:1992wi}   & Ref.\cite{Ivanov:1997ra} &Ref.\cite{Zenczykowski:1993jm}& Ref.\cite{Xu:1992vc}  &  Ref.\cite{Cheng:1991sn}&Ref.\cite{Cheng:1993gf}&Ref.\cite{Sharma:1998rd}&Exp\cite{PDG18}\\\hline
 $\Gamma$ &1.10 &3.03  &1.34& 1.17     &2.48 &6.07&4.62&$4.08\pm0.33$ \\\hline
 $\alpha$ & 0.70 &0.43  &0.39& 0.92  &0.83&-0.49&-0.31&$-0.45\pm0.32$
\end{tabular}
\end{ruledtabular}
\end{table}

\section{The sequential decay  $\Lambda_c\to\Lambda\rho\to \Sigma \pi$}

The amplitude of  baryon decays $\mathcal{B}_i\to \mathcal{B}_f P$ can be written as\cite{Cheng:1991sn}
\begin{eqnarray}
\mathcal{M}(\mathcal{B}_i\to \mathcal{B}_f P)=i\bar U_{\mathcal{B}_f }[A-B\gamma_5] U_{\mathcal{B}_i},
\end{eqnarray}
where $\mathcal{B}_i$ ($\mathcal{B}_f$) is the initial (final) baryon and $P$ is a pseudoscalar meson.

For the transition $\mathcal{B}_i\to \mathcal{B}_f V$, the amplitude is
\begin{eqnarray}
\mathcal{M}(\mathcal{B}_i\to \mathcal{B}_f V)=i\bar U_{\mathcal{B}_f }\varepsilon^{*\mu}[-A_1\gamma_\mu\gamma_5-A_2p_{f\mu}\gamma_5+B_1\gamma_\mu+B_2p_{f\mu}] U_{\mathcal{B}_i},
\end{eqnarray}
where $V$ is a vector meson with polarization $\varepsilon$ and $p_f$ is the momentum of $\mathcal{B}_f$. It is noted that the sign of the $\gamma_5$ term
in Ref.\cite{Cheng:1991sn} is opposite to the convention adopted in \cite{Cheng:2018hwl}. Here with this minus sign in front of every item involving $\gamma_5$ (conventional definition), all the values of $B$, $A_1$ and $A_2$ given in Ref.\cite{Cheng:1991sn} do not need to be  changed and we  directly adopt their formulas. Indeed. this provides us a great convenience to derive relevant quantities.

For the decay $\Lambda_c\to \Sigma\pi$ the factors $A$ and $B$ should include the contributions of all relevant Feynman diagrams. At the quark level the transition does not occur
via factorizable  Feynman diagrams but those of non-factorizable ones\cite{Korner:1992wi,Cheng:1991sn}. In Ref.\cite{Cheng:1991sn} the authors employed the simple pole-model to calculate the contributions of non-factorizable  Feynman diagrams. However the sign of the up-down asymmetry $\alpha$ gained in this way is opposite to data.

It is easy to conjecture that in this case, the contribution from the re-scattering of final state in some decays of $\Lambda_c$ might play an important role. The goal of this
work is just to check if the re-scattering of the final products can change the scenario, namely simultaneously results in the required production rate for $\Lambda_c\to\Sigma\pi$ and a correct up-down asymmetry $\alpha$.
In principle  many coupled channels would jointly contribute to the decay $\Lambda_c\to\Sigma\pi$, for example $\Lambda_c\to \Lambda\rho$ and $\Lambda_c\to \Lambda\pi$ etc.
Considering the coupling constants and the rates of the first step decay of $\Lambda_c$ one can decide that the main process is $\Lambda_c\to \Lambda\rho\to\Sigma\pi$
where the second step is the isospin conserving re-scattering $\Lambda\rho\to\Sigma\pi$.

The total amplitude of the practical transition $\Lambda_c\to \Sigma\pi$ is
\begin{eqnarray}
M_{(\Lambda_c\to\Sigma\pi)}&&= M^{\rm DIR}_{(\Lambda_c\to\Sigma\pi)}+M^{\rm FSI}_{(\Lambda_c\to\Lambda\rho\to \Sigma\pi)}\nonumber\\
&&=i\bar U_{\Sigma}[A^{\rm DIR}-B^{\rm DIR}\gamma_5] U_{\Lambda_c}+i\bar U_{\Sigma}[A^{\rm FSI}-B^{\rm FSI}\gamma_5] U_{\Lambda_c}\nonumber\\
&&=i\bar U_{\Sigma}[A-B\gamma_5] U_{\Lambda_c},
\end{eqnarray}
where $M^{\rm DIR}_{(\Lambda_c\to\Sigma\pi)}$ and $M^{\rm FSI}_{(\Lambda_c\to\Lambda\rho\to \Sigma\pi)}$
correspond to the contributions of the direct transition $\Lambda_c\to\Sigma\pi$ and the two-step process $\Lambda_c\to\Lambda\rho\to \Sigma\pi$ respectively, $A=A^{\rm DIR}+A^{\rm FSI}$ and $B=B^{\rm DIR}+B^{\rm FSI}$.
The amplitude of the direct transition $M^{\rm DIR}_{(\Lambda_c\to\Sigma\pi)}$ was calculated in terms of the pole model \cite{Cheng:1991sn} and we will use their numerical results directly.

\begin{figure*}
        \centering
        \subfigure[~]{
          \includegraphics[width=7cm]{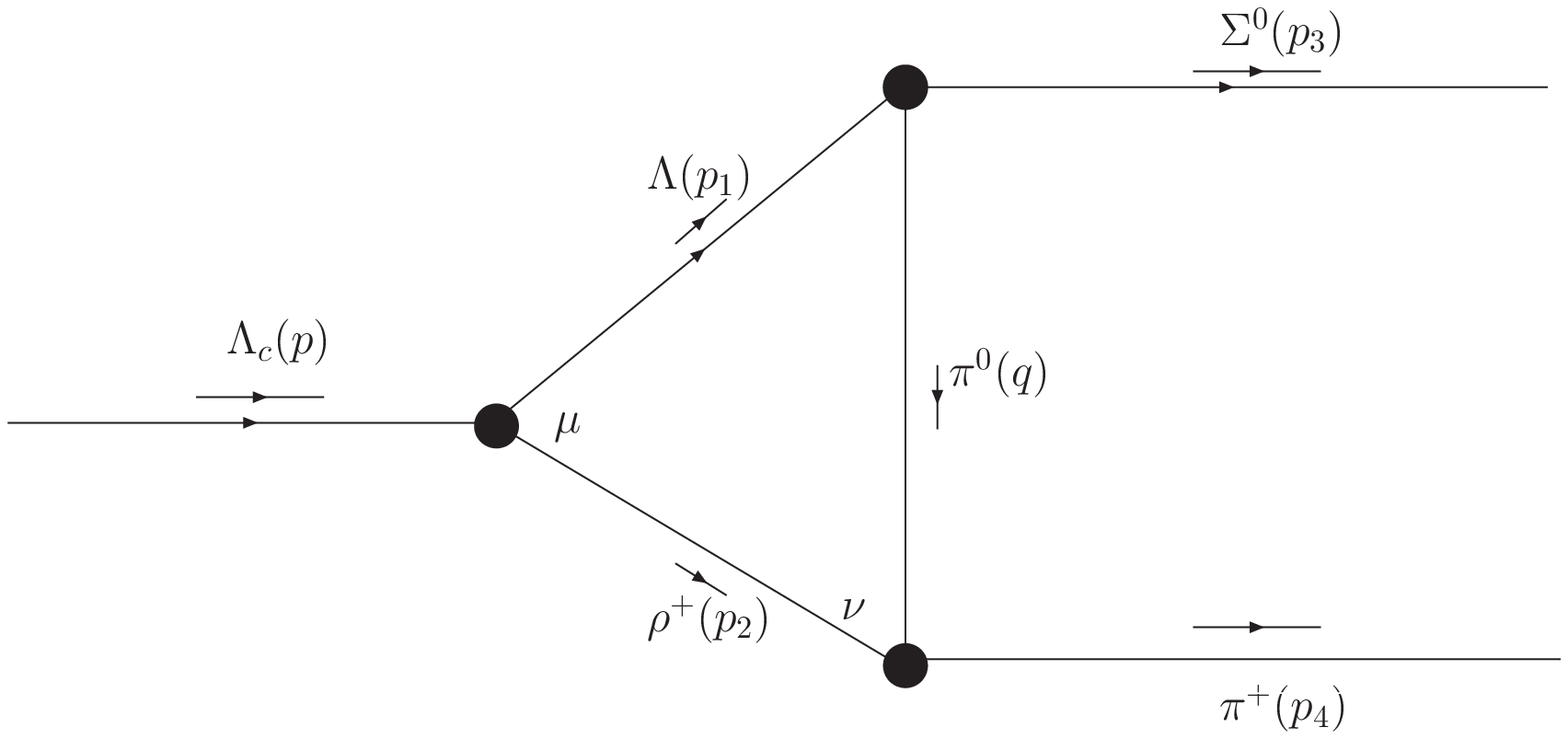}}\,\,\,\,\,\,\,\,\,\,\,\,\,\,\,\,\,\,\,\,\,\,
        \subfigure[~]{
          \includegraphics[width=7cm]{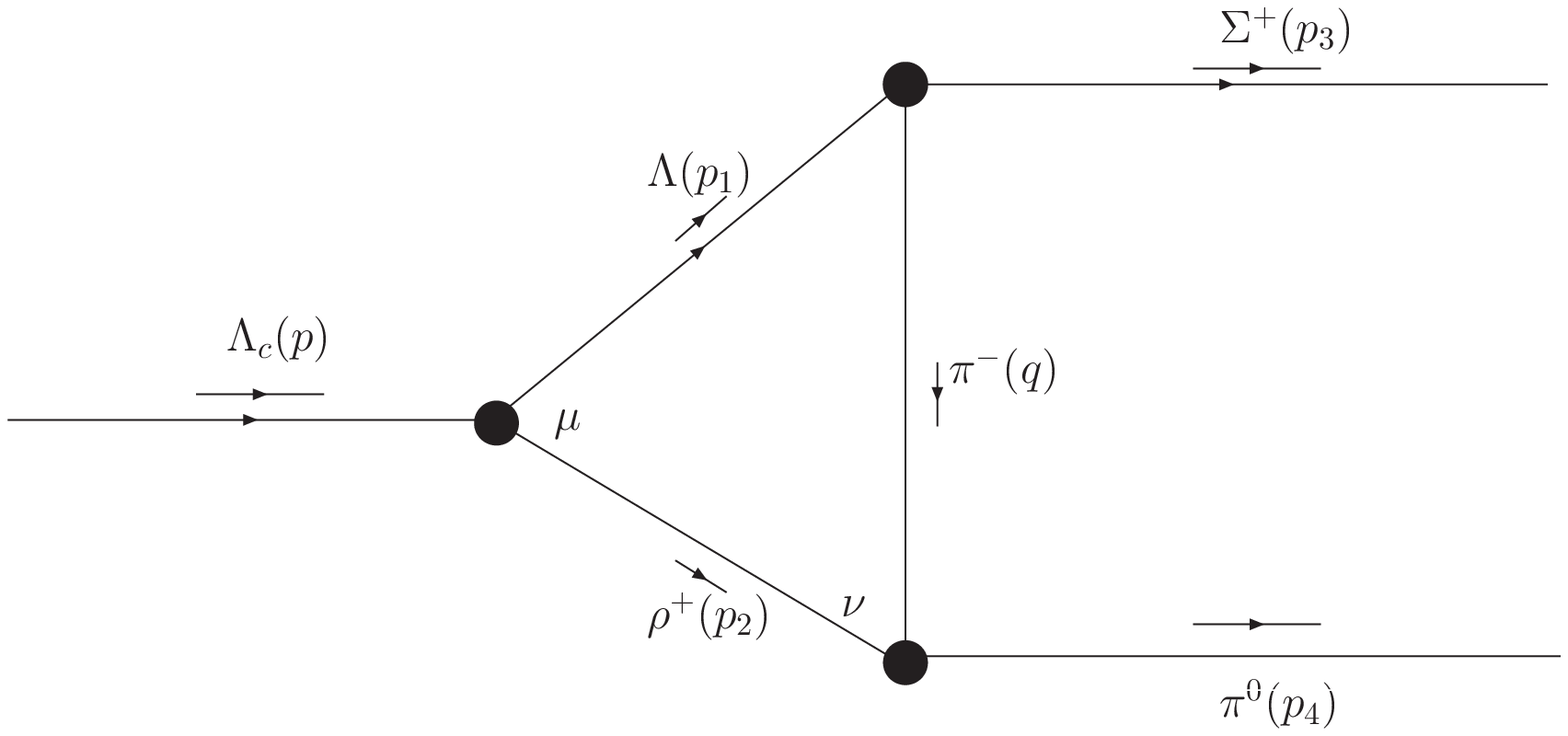}}
 \caption{The main final state interaction for $\Lambda_c\to \Sigma^0 \pi^+$ (a) and $\Lambda_c\to \Sigma^+ \pi^0$ (b).}
        \label{t1a}
    \end{figure*}

Now let us begin to study the processes whose corresponding Feynmen diagrams are depicted in Fig.\ref{t1a}.

The relevant effective interactions are
\cite{Shen:2016tzq}
\begin{eqnarray}
&&\mathcal{L}_{_{\Lambda\Sigma\pi}}=ig_{_{\Lambda\Sigma\pi}}\bar
\psi_{_{\Sigma}}\gamma_5\psi_{_{\Lambda}}\pi\nonumber\\\mathcal{L}_{\rho\pi\pi}=&&g_{_{\rho\pi\pi}}(\partial_\mu \pi^0 \pi^+\rho^{-\mu}-\partial_\mu
\pi^+\pi^0\rho^{-\mu} ),
\end{eqnarray}

Generally, the absorptive part overwhelmingly dominates and the contribution of
 the dispersive one can be ignored, so that we only need to calculate the absorptive part of the Feynmen diagrams in Fig.\ref{t1a}.
where the intermediate states $\rho$ and $\Lambda$ are on shell and the transition amplitude of $\Lambda_c\to\Lambda\rho$ can also
be written in terms of the pole model. By the Cutkosky rule, one can factorize the transition into two parts as
$M(\Lambda\to\Lambda\rho)M(\Lambda\rho\to \Sigma\pi)$. Let us first
calculate the amplitude corresponding to diagram Fig 1 (a):

\begin{eqnarray}
M^{\rm FSI}_{(\Lambda_c\to\Lambda\rho^+\to\Sigma^0\pi^+)}=&&\frac{1}{2}\int
\frac{d\mathbf{p}_1}{(2\pi)^32E_1}\frac{d\mathbf{p}_2}{(2\pi)^32E_2}(2\pi)^4\delta(p-p_1-p_2)\mathcal{M}[\Lambda_c\to\Lambda\rho]\mathcal{M}[\Lambda\rho\to\Sigma
\pi]\nonumber\\=&& \frac{1}{2}\int
\frac{d\mathbf{p}_1}{(2\pi)^32E_1}\frac{d\mathbf{p}_2}{(2\pi)^32E_2}(2\pi)^4\delta(p-p_1-p_2)i\bar
U_{\Lambda}[-A_1\gamma_\mu\gamma_5-A_2p_{1\mu}\gamma_5\gamma_\mu\nonumber\\&&+B_1+B_2p_{1\mu}]
U_{\Lambda_c} g_{_{\Sigma\Lambda\pi}}\bar
U_{\Sigma}i\gamma_5
U_{\Lambda}g_{_{\rho\pi\pi}}i(p_4+q)_\nu(-g^{\mu\nu}+\frac{p_2^\mu
p_2^\nu}{m_\rho^2})\frac{i}{q^2
-m^2_\pi}F^2(q^2)\nonumber\\=&&\int
\frac{|\mathbf{p}_1|d\Omega}{32\pi^2E}\bar
U_{\Lambda}[-A_1\gamma_\mu\gamma_5-A_2p_{1\mu}\gamma_5+B_1\gamma_\mu+B_2p_{1\mu}]
U_{\Lambda_c}\bar U_{\Sigma}\gamma_5 U_{\Lambda}\nonumber\\&&
g_{_{\Sigma\Lambda\pi}}g_{_{\rho\pi\pi}}(p_4+q)_\nu(-g^{\mu\nu}+\frac{p_2^\mu
p_2^\nu}{m_\rho^2})\frac{F^2(q^2)}{q^2-m^2_\pi}\nonumber\\=&&\int
\frac{|\mathbf{p}_1|d\Omega}{32\pi^2E}\bar U_{\Sigma}\gamma_5
(p_1\!\!\!\!\slash+m_\Lambda)[-A_1\gamma_\mu\gamma_5-A_2p_{1\mu}\gamma_5+B_1\gamma_\mu+B_2p_{1\mu}]
U_{\Lambda_c}\nonumber\\&&
g_{_{\Sigma\Lambda\pi}}g_{_{\rho\pi\pi}}(p_4+q)_\nu(-g^{\mu\nu}+\frac{p_2^\mu
p_2^\nu}{m_\rho^2})\frac{F^2(q^2)}{q^2-m^2_\pi},
\end{eqnarray}
where $E$, $E_1$ and $E_2$ are the energies of $\Lambda_c$, $\Lambda$ and $\rho$, with $p$, $p_1$, $p_2$, $p_3$ and $p_4$ being the momenta of  $\Lambda_c$, $\Lambda$, $\rho$, $\Sigma$ and $\pi$, $q$ is the momentum of the exchanged intermediate pions. Since in practice meson and baryon are not point particles,  a form
factor at each effective vertex should be introduced. The form factor
suggested by many researchers  is in the form:
\begin{eqnarray} \label{form-factor} F(q,m_{ P}^2 ) = {\Lambda_1^2 -
m_{P}^2 \over \Lambda_1^2 -{q}^2}\,,
\end{eqnarray}
where $\Lambda_1$ is a cutoff parameter and $m_{ P}$ is equal to $m_{\pi}$. Since the form factor is
not derived from a fundamental principle, and the concerned cutoff
parameter is neither determined theoretically, actually so far we
know little about the cutoff parameter $\Lambda_1$. In some
Refs.\cite{Cheng:2004ru,Ke:2010aw} the form
factor is parameterized as $\Lambda_1=\lambda\Lambda_{QCD}+m_P$ with
$\Lambda_{QCD}\approx 220$ MeV and the dimensionless parameter $\lambda$
is of order of unit.

Using the four-momentum relations $p_4=p-p_3$, $q=p_1-p_3$, $p_2=p-p_1$ and contracting the indices $\mu$ and $\nu$, these notations $p\!\!\slash$, $p_1\!\!\!\!\slash$, $p_3\!\!\!\!\slash$
, $p^2$, $p_1^2$, $p_3^2$, $p\cdot p_1$, $p\cdot p_3$ and $p_1\cdot p_3$ appear in the expression of $M^{\rm FSI}_{(\Lambda_c\to\Lambda\rho^+\to\Sigma^0\pi^+)}$. One can employ Dirac equations $p\!\!\slash U_{\Lambda_c}=mU_{\Lambda_c}$ and $p_3\!\!\!\!\slash U_{\Sigma}=m_3U_{\Sigma}$ to simply the expression. Since $\Lambda_c$, $\Sigma$ and $\Lambda$ are on-shell, $p^2$, $p_1^2$, $p_3^2$, $p\cdot p_1$, $p\cdot p_3$ and $p_1\cdot p_3$ can be expressed in terms of observable physical quantities. At last, one needs to deal with $p_1\!\!\!\!\slash$. In our calculation we choose $\mathbf{p}_3$  in  $z-$direction and the
angle spanned between $\mathbf{p}_1$ and $\mathbf{p}_3$ is $\theta$. Since there exists an integration over azimuth one can find $p_1\!\!\!\!\slash=C_1p\!\!\slash+C_2p_3\!\!\!\!\slash$ with $C_1=\frac{E_1|\mathbf{p}_3|-|\mathbf{p}_1|E_3{\rm cos}\theta}{m|\mathbf{p}_3|}$ and $C_2=\frac{|\mathbf{p}_1|{\rm cos}\theta}{|\mathbf{p}_3|}$.
Finally we obtain
\begin{eqnarray}
M^{\rm FSI}_{(\Lambda_c\to\Lambda\rho^+\to\Sigma^0\pi^+)}=i\bar U_{\Sigma}[A^{\rm FSI}-B^{\rm FSI}\gamma_5] U_{\Lambda_c},
\end{eqnarray}
with
$
{A^{\rm FSI}}=g_{_{\Sigma\Lambda\pi}}g_{_{\rho\pi\pi}}\int \frac{C_a|\mathbf{p}_1|{\rm sin} \theta F^2(q^2)
d\theta}{16\pi E}$ and ${B^{\rm FSI}}=g_{_{\Sigma\Lambda\pi}}g_{_{\rho\pi\pi}}\int
\frac{C_b|\mathbf{p}_1|{\rm sin} \theta F^2(q^2) d\theta}{16\pi E}$.
The detailed expressions of $C_a$ and $C_b$ are
\begin{eqnarray}
C_a=&&{-A_1}\,[ m^3\,{m_1} + m^2\,{{m_1}}^2 - {{m_1}}^4 -
2\,{E_3}\,m\,{m_1}\,( m + {m_1} )  +
       {{m_1}}^2\,{{m_2}}^2 + 2\,{m_1}\,{{m_2}}^2\,{m_3}\nonumber\\&& -
       m\,{m_1}\,( {{m_1}}^2 + {{m_2}}^2 - 2\,{p_1\cdot p_3} )  + 2\,{{m_1}}^2\,{p_1\cdot p_3} -
       4\,{{m_2}}^2\,{p_1\cdot p_3} ]/[{{m_2}}^2\,( {{m_{\pi}}}^2 - {q^2} ) \nonumber\\&& - {A_2}\,{m_1}\,
     [ -2\,{E_3}\,m\,{{m_1}}^2 + m^2\,{{m_1}}^2 - {{m_1}}^4 + {{m_1}}^2\,{{m_2}}^2 +
       {E_1}\,m\,( 2\,{E_3}\,m - m^2 + {{m_1}}^2 \nonumber\\&&+ {{m_2}}^2 - 2\,{p_1\cdot p_3} )  +
       2\,{{m_1}}^2\,{p_1\cdot p_3} - 2\,{{m_2}}^2\,{p_1\cdot p_3} ] /[{{m_2}}^2\,( {{m_{\pi}}}^2 - {q^2} )
       ]\nonumber\\
&&-\,{(m C_1+m_3 C_2)} {A_1}\,[ -m^3 - m^2\,{m_1} + {{m_1}}^3 +
2\,{E_3}\,m\,( m + {m_1} )  -
         {m_1}\,{{m_2}}^2 + 2\,{{m_2}}^2\,{m_3} \nonumber\\&&+ m\,( {{m_1}}^2 + {{m_2}}^2 - 2\,{p_1\cdot p_3} )  -
         2\,{m_1}\,{p_1\cdot p_3} ]/[{{m_2}}^2\,
    ( {{m_{\pi}}}^2 - {q^2} ) ] \nonumber\\&& + {(m c_1+m_3 c_2)}{A_2}\,
       [ -2\,{E_3}\,m\,{{m_1}}^2 + m^2\,{{m_1}}^2 - {{m_1}}^4 + {{m_1}}^2\,{{m_2}}^2 +
         {E_1}\,m\,( 2\,{E_3}\,m \nonumber\\&&- m^2 + {{m_1}}^2 + {{m_2}}^2 - 2\,{p_1\cdot p_3} )  +
         2\,{{m_1}}^2\,{p_1\cdot p_3} - 2\,{{m_2}}^2\,{p_1\cdot p_3} ]   /[{{m_2}}^2\,
    ( {{m_{\pi}}}^2 - {q^2} ) ],\nonumber\\
C_b=&&{B_1}  [2  {E_3} m  {m_1} (m- {m_1})-m^3  {m_1}+m^2  {m_1}^2+m
{m_1}  ( {m_1}^2+ {m_2}^2-2
    {p_1\cdot p_3} )- {m_1}^4+ \nonumber\\&&{m_1}^2  {m_2}^2+2  {m_1}^2  {p_1\cdot p_3}+2  {m_1}  {m_2}^2  {m_3}-4  {m_2}^2  {p_1\cdot p_3} ]/[ {m_2}^2  ( {m_{\pi}}^2- {q^2} )]\nonumber\\&&+ {B_2}
    {m_1}  [ {E_1} m  (2  {E_3} m-m^2+ {m_1}^2+ {m_2}^2-2  {p_1\cdot p_3} )-2  {E_3} m  {m_1}^2+m^2  {m_1}^2- {m_1}^4\nonumber\\&&+ {m_1}^2
    {m_2}^2+2  {m_1}^2  {p_1\cdot p_3}-2  {m_2}^2  {p_1\cdot p_3}]/[ {m_2}^2  ( {m_{\pi}}^2- {q^2} )]
\nonumber\\&&+ {(m c_1-m_3 c_2)}{B_1}  [2  {E_3} m (m-
{m_1})-m^3+m^2 {m_1}+m ( {m_1}^2+ {m_2}^2-2  {p_1\cdot p_3}
)\nonumber\\&&- {m_1}^3+ {m_1}
    {m_2}^2+2  {m_1}  {p_1\cdot p_3}-2  {m_2}^2  {m_3} ]/[ {m_2}^2
    ( {m_{\pi}}^2- {q^2} )]\nonumber\\&&+(m C_1-m_3 C_2){B_2}  [ {E_1} m  (2  {E_3} m-m^2+ {m_1}^2+ {m_2}^2-2  {p_1\cdot p_3} )-2
    {E_3} m  {m_1}^2+m^2  {m_1}^2\nonumber\\&&- {m_1}^4+ {m_1}^2  {m_2}^2+2  {m_1}^2  {p_1\cdot p_3}-2  {m_2}^2  {p_1\cdot p_3} ] /[ {m_2}^2
    ( {m_{\pi}}^2- {q^2} )],
\end{eqnarray}
where $m$, $m_1$, $m_2$, $m_3$ and $m_4$ are the masses of $\Lambda_c$, $\Lambda$, $\rho$, $\Sigma$ and $\pi$ respectively, $E_3$ is the energy of $\Sigma$.

By the same process one obtains the absorptive part of the Feynman diagram (b) on the right panel of Fig.\ref{t1a}
and it is noted that
\begin{eqnarray}
M^{\rm FSI}_{(\Lambda_c\to\Lambda\rho^+\to\Sigma^+\pi^0)}=-M^{\rm FSI}_{(\Lambda_c\to\Lambda\rho^+\to\Sigma^0\pi^+)},
\end{eqnarray}
where the minus sign comes from the effective interaction $\mathcal{L}_{_{\rho\pi\pi}}$ i.e. the sign of the absorptive part of $\Lambda_c\to\Sigma^0\pi^+$ is opposite to that of $\Lambda_c\to\Sigma^+\pi^0$. The minus sign is necessary for determining the symbol of the asymmetry parameter of the decay $\Lambda_c\to\Sigma^+\pi^0$.

The decay rates of $\mathcal{B}_i\rightarrow \mathcal{B}_fP$
and up-down asymmetries are \cite{Cheng:1991sn}
 \begin{eqnarray}
 \Gamma&=&\frac{|\mathbf{p}_c|}{8\pi}\left[\frac{(m_i+m_f)^2-m_P^2}{m_i^2}|A|^2+
  \frac{(m_i-m_f)^2-m_P^2}{m_i^2}|B|^2\right], \non\\
 \alpha&=&\frac{2\kappa{\rm Re}(A^*B)}{|A|^2+\kappa^2|B|^2},
 \end{eqnarray}
where $\mathbf{p}_c$ is the three-momentum of $\mathcal{B}_f$ in the rest frame of
$\mathcal{B}_i$ and $\kappa=\frac{ |\mathbf{p}_c|}{E_{f}+m_f}$.

The amplitude of the direct transition $\Lambda_c\to\Sigma\pi$ is straightforward calculated in terms of the pole model which was shown in
the early works, so that we omit the details of the calculations in this section.

\section{The theoretical predictions as the contribution from final state interaction is taken into account}

In a new work, the authors restudied the Cabibbo-suppressed decays of $\Lambda_c$ \cite{Cheng:2018hwl} where the 30 years-old work
\cite{Cheng:1991sn,Cheng:1993gf} was cited. It indicates, the fundamental framework does not change and therefore we can employ their numerical results directly. Moreover,
$\Lambda_c\to \Lambda\rho$ has also been investigated
\cite{Cheng:1991sn}, thus we just directly use their results about the first step transitions $M(\Lambda_c\to\Lambda\rho)$. Then we
concentrate our attention on exploring the
re-scattering process $\Lambda\rho\to \Sigma\pi$ and analyze the consequences especially how inclusion of the final state interaction  results in
an opposite sign for the up-down asymmetry parameter from that determined by the direct transition.

\begin{table}
\caption{Theoretical results of $\Lambda_c\to \Sigma \pi$ in pole model where $A^{\rm fac}$, $A^{\rm pole}$, $B^{\rm fac}$, $B^{\rm pole}$ are in units of $G_FV_{cs}V_{ud}\times 10^{-2}$GeV$^2$ and $\Gamma$ is in units of $10^{-14}$GeV}\label{Tab:11}
\begin{ruledtabular}
\begin{tabular}{cccccccc}
  &$A^{\rm fac}$ & $A^{\rm pole}$  &  $B^{\rm fac}$   &  $B^{\rm pole}$ &$\alpha$&$\Gamma$\\\hline
 $\Lambda_c\to \Sigma^0 \pi^+$ & 0 & 2.24     &0 &14.63  &0.83&2.48 \\\hline
 $\Lambda_c\to \Sigma^+ \pi^0$ & 0 & -2.24     &0 &-14.63  &0.83&2.48
\end{tabular}
\end{ruledtabular}
\end{table}

\begin{table}
\caption{Theoretical results of $\Lambda_c\to \Lambda\rho$ in pole model where $A_1^{\rm fac}$, $A_1^{\rm pole}$, $A_2^{\rm fac}$, $A_2^{\rm pole}$, $B_1^{\rm fac}$, $B_1^{\rm pole}$, $B_2^{\rm fac}$, $B_2^{\rm pole}$ are in units of $G_FV_{cs}V_{ud}\times 10^{-2}$GeV$^2$ and $\Gamma$ is in units of $10^{-14}$GeV}\label{Tab:12}
\begin{ruledtabular}
\begin{tabular}{ccccccccccc}
 $\Lambda_c\to  \Lambda\rho^+$ &$A_1^{\rm fac}$ & $A_1^{\rm pole}$   &$A_2^{\rm fac}$ & $A_2^{\rm pole}$ &  $B_1^{\rm fac}$   &  $B_1^{\rm pole}$ &  $B_2^{\rm fac}$   &  $B_2^{\rm pole}$  &$\alpha$&$\Gamma$\\\hline
(a)& -8.64 &0     &-0.71 &0  & 13.33&0&-2.99&0&-0.30&14.13\\\hline
(b) & -8.28 &0     &-0.68 &0  & 12.77&0&-2.87&0&-0.30&12.97
\end{tabular}
\end{ruledtabular}
\end{table}

We list the theoretical predictions given by the authors of Ref.\cite{Cheng:1991sn} in Tab. \ref{Tab:11} and \ref{Tab:12} for a clear reference.
It is noted that the authors corrected their values about $\Lambda_c\to  \Lambda\rho$ later and we employ the new ones for  our numerical computations
(line (a) in Tab. \ref{Tab:12}). Since the effective color-favored Wilson coefficient ($\sim 1.315$) used in Ref.\cite{Cheng:1991sn} is larger than the present values ($\sim 1.26$)\cite{Cheng:2018hwl} we set it to be 1.26 and repeat the calculations. The values we obtained are listed in line (b) of Tab. \ref{Tab:12} .

\begin{table}
\caption{ $A^{\rm FSI}$ and $B^{\rm FSI}$ in  $M^{\rm FSI}_{\Lambda_c\to \Sigma \pi}$  ( in units of $G_FV_{cs}V_{ud}\times 10^{-2}$GeV$^2$ )}\label{Tab:13}
\begin{ruledtabular}
\begin{tabular}{cccccc}
  &$A^{\rm FSI}$ &  $B^{\rm FSI}$  \\\hline
 $\Lambda_c\to \Sigma^0 \pi^+ \,(\Lambda_1=0.8$ GeV)  & -3.26 (-3.12)    &2.75 (2.63)   \\\hline
 $\Lambda_c\to \Sigma^+ \pi^0\,(\Lambda_1=0.8$ GeV)    &3.26 (3.12)    &-2.75 (-2.63) \\\hline
 $\Lambda_c\to \Sigma^0 \pi^+ \,(\Lambda_1=1$ GeV) & -4.68 (-4.48)    &5.17 (4.95) \\\hline
 $\Lambda_c\to \Sigma^+ \pi^0\,(\Lambda_1=1$ GeV)&4.68 (4.48)    &-5.17 (-4.95)
\end{tabular}
\end{ruledtabular}
\end{table}

\begin{table}
\caption{Theoretical total results of $\Lambda_c\to \Sigma \pi$ where $A$ and $B$ are in units of $G_FV_{cs}V_{ud}\times 10^{-2}$GeV$^2$ and $\Gamma$ is in units of $10^{-14}$GeV}\label{Tab:14}
\begin{ruledtabular}
\begin{tabular}{cccccccc}
  &$A$ &  $B$   &$\alpha$&$\Gamma$\\\hline
 $\Lambda_c\to \Sigma^0 \pi^+ \,(\Lambda_1=0.8$ GeV)& -1.02 (-0.88)    &17.38 (17.26) &-0.29 (-0.25)&2.95 (2.89) \\\hline
 $\Lambda_c\to \Sigma^+ \pi^0\,(\Lambda_1=0.8$ GeV)&1.02 (0.88)    &-17.38 (-17.26)  &-0.29 (-0.25)&2.95 (2.89)\\\hline
 $\Lambda_c\to \Sigma^0 \pi^+ \,(\Lambda_1=1$ GeV)  & -2.44 (-2.24)    &19.80 (19.58)  &-0.64 (-0.60)&4.24 (4.07) \\\hline
 $\Lambda_c\to \Sigma^+ \pi^0\,(\Lambda_1=1$ GeV)  &2.44 (2.24)    &-19.80 (-19.58)   &-0.64 (-0.60)&4.24 (4.07)
\end{tabular}
\end{ruledtabular}
\end{table}

In order to perform the numerical computations we need to determine the coupling constant $g_{_{\Sigma\Lambda\pi}}$ and $g_{_{\rho\pi\pi}}$. Using the data in
particle data book\cite{PDG18} we fix $g_{_{\rho\pi\pi}}=6.01$ and another factor $g_{_{\Sigma\Lambda\pi}}=11.8$ was given in Ref.\cite{Cheng:1991sn}.
Generally the cut-off parameter $\Lambda_1$ is about 1 GeV for a light exchanged meson. In our calculation we set it to be 0.8 GeV and 1 GeV respectively to make more sense.
Using the formula derived above $A^{\rm FSI}$ and $B^{\rm FSI}$ are calculated with $A_1=A^{\rm fac}_1+A^{\rm pole}_1$, $A_2=A^{\rm fac}_2+A^{\rm pole}_2$, $B_1=B^{\rm fac}_1+B^{\rm pole}_1$ and $B_2=B^{\rm fac}_2+B^{\rm pole}_2$ and their numerical values are presented in Tab.\ref{Tab:13}.
Using the values in table \ref{Tab:11} one can obtain $A^{\rm DIR}=A^{\rm fac}+A^{\rm pole}$ and $B^{\rm DIR}=B^{\rm fac}+B^{\rm pole}$.
Summing up the contributions of the direct transition and that involving final state interaction we have $A$ and $B$ in the total amplitude $M_{(\Lambda_c\to\Sigma\pi)}$
and our theoretical results are presented in Tab. \ref{Tab:13} where the values in front of (or between) the parentheses are corresponding to those in Tab. \ref{Tab:12} (a) (or (b)). The experimental results on $\Gamma(\Lambda_c\to \Sigma^+ \pi^0)$ and $\alpha$ are $(4.08\pm0.33)\times10^{-14}$GeV and $-0.45\pm0.32$.

In this scenario, we have made a theoretical prediction on the up-down asymmetry $\alpha$ whose  value
resides within the error tolerance of the data and its sign is consistent with the experimental measurement.
The predicted decay width of $\Gamma(\Lambda_c\to \Sigma \pi)$  is also closer to data than that made
in Ref.\cite{Cheng:1991sn}. Apparently final state interaction changes the naive results of Ref.\cite{Cheng:1991sn}. From table \ref{Tab:11} one can find that the signs of $A^{\rm pole}$ and $B^{\rm pole}$ are the same so the sign of $\alpha$ calculated in pole model is positive. As  the final state interaction are taken into account
the interference between the direct transition and the final state interaction would induce a conversion of the sign of the asymmetry parameter. Namely
$A$ and $B$ (table \ref{Tab:14}) possess  opposite signs, so the sign of $\alpha$ in is negative.
  It is noticed that in this work, only the contribution from the decay $\Lambda_c\to  \Lambda\rho$ as an intermediate state is accounted, certainly in principle,
other decay portals of $\Lambda_c$
should also contribute to the same process via re-scattering. A careful analysis indicates that those contributions are not as important as that of
$\Lambda_c\to  \Lambda\rho\to\Sigma\pi$, therefore, we ignore those coupled channels in this work.

\section{summary}
At the quark level the decay $\Lambda_c\to \Sigma\pi$ receives only the non-factorizable $W$-exchange and internal $W$-emission contributions. Based on the valence quark model\cite{Korner:1992wi,Ivanov:1997ra} or the pole-model\cite{Xu:1992vc,Cheng:1991sn,Zenczykowski:1993jm} these  non-factorizable diagrams were calculated while the resultant sign of the up-down asymmetry conflicts with data. Employing current algebra the authors\cite{Sharma:1998rd,Cheng:1995fe} obtained a negative up-down asymmetry as required by data. Generally current algebra can be applied to study the decays where a soft pseudoscalar meson is emitted. However the pion in $\Lambda_c\to \Sigma\pi$ is far from being soft so it is not natural to explain the data by using the current algebra. Following the approach in the references\cite{Meng:2008dd,Ke:2010aw,Cheng:2004ru,Yuan:2012zw} we suggest that a final state interaction (or re-scattering) in the decays of $\Lambda_c$ can contribute to the observed $\Lambda_c\to \Sigma\pi$. In terms of the effective interactions, coupling constants
we calculate the contribution of the subprocess $\Lambda_c\to\Lambda\rho\to \Sigma\pi$ to the observed $\Lambda_c\to\Sigma\pi$. We notice (see the Table \ref{Tab:11}), with the pole model the contribution to both $A^{\rm DIR}$ and $B^{\rm DIR}$ in
Eq.(1) are positive while the the contributions originating from the re-scattering to $A^{\rm FSI}$ destructively interferes with that of  $A^{\rm DIR}$  of the direct transition whereas
$B^{\rm FSI}$ constructively interferes with  $B^{\rm DIR}$, thus as a consequence, the sign of the asymmetry parameter is reversed due to the destructive interference.

For our concrete calculations, generally considering, if there exists an absorptive part, it should dominate the rate. Thus we only calculate the absorptive part of the triangle (see the Feynman diagrams) where in the intermediate step
$\Lambda$ and $\rho$ are on-shell and we can factorize the two steps $\Lambda_c\to\Lambda\rho$ and $\Lambda\rho\to\Sigma\pi$. Including the contribution of the direct transition $\Lambda_c\to \Sigma\pi$ calculated using pole model and the subprocess $\Lambda_c\to\Lambda\rho\to\Sigma\pi$ which involves re-scattering effects, we obtain a negative up-down asymmetry $\alpha$ and the resultant decay width of $\Gamma(\Lambda_c\to \Sigma \pi)$ is also closer to data than the original results of Ref.\cite{Cheng:1991sn}. It is also noted that the re-scattering of
$\Lambda\rho$ to other products may reduce the observed  rate of $\Lambda_c\to\Lambda\rho$. We predict the observed width of $\Lambda_c\to\Lambda\rho$ would be somehow smaller than the theoretically predicted value given in Tab. \ref{Tab:12}.

In this paper we  study the contribution of final state interaction to the transition $\Lambda_c\to \Sigma\pi$ and confirm that the final state interaction plays an important role in many hadronic transitions. In fact there still exist some discrepancies between theoretical estimations and data for other decays of $\Lambda_c$, and we hope the mechanism can also be applied to studying those ``anomalies".

\section*{Acknowledgement}
This work is supported by the National Natural Science Foundation
of China (NNSFC) under Contract No. 11375128, 11675082 and 11735010.

\end{document}